\documentclass[twocolumn,preprintnumbers,amsmath,amssymb,superscriptaddress]{revtex4}
\usepackage{graphicx}
\usepackage{epstopdf}
\usepackage{dcolumn}
\usepackage{bm}

\begin{document}

\title{Laser cooling of a Planck mass object close to the quantum ground state}

\author{L. Neuhaus}
\author{R. Metzdorff}
\author{S. Zerkani}
\author{S. Chua}
\author{J. Teissier\footnote{Current Affiliation: II-VI Laser Enterprise, Binstrasse 17, CH-8045 Zurich, Switzerland}}
\author{D. Garcia-Sanchez\footnote{Current affiliation: Institut des NanoSciences de Paris, Sorbonne Universit\'e, CNRS-UMR 7588,  F-75005, Paris, France}}
\author{S. Del\'eglise}
\author{T. Jacqmin}
\author{T. Briant}
\affiliation{Laboratoire Kastler Brossel, Sorbonne Universit\'e, ENS - Universit\'e PSL, Coll\`ege de France, CNRS, Paris,
France}
\author{J. Degallaix}

\author{V. Dolique\footnote{Current Affiliation: Univ Lyon, ENS de Lyon, Univ Claude Bernard Lyon 1, CNRS, Laboratoire de Physique, F-69342 Lyon, \author{C. Michel}
\author{L. Pinard}\affiliation{Universit\'e de Lyon, Universit\'e Claude Bernard Lyon 1, CNRS, Laboratoire des Mat\'eriaux Avanc\'es (LMA), IP2I Lyon / IN2P3, UMR 5822, F-69622 Villeurbanne, France}
France}}
\author{G. Cagnoli}
\affiliation{Universit\'e de Lyon, Universit\'e Claude Bernard Lyon 1, CNRS, Institut Lumi\`ere
Mati\`ere, F-69622, Villeurbanne, France}
\author{O. Le Traon}
\author{C. Chartier}
\affiliation{ONERA—The French Aerospace Lab, F-91123 Palaiseau Cedex, France}
\author{A. Heidmann}
\author{P.-F. Cohadon}
\email{cohadon@lkb.upmc.fr}
\affiliation{Laboratoire Kastler Brossel, Sorbonne Universit\'e, ENS - Universit\'e PSL, Coll\`ege de France, CNRS, Paris,
France}

\begin{abstract}
Quantum mechanics has so far not been tested for mechanical objects at the scale of the Planck mass $\sqrt{\hbar c/ G} \simeq 22\,\mu\mathrm{g}$. We present an experiment where a 1 mm quartz micropillar resonating at 3.6 MHz with an effective mass of 30 $\mu$g is cooled to 500 mK with a dilution refrigerator, and further optomechanically sideband-cooled to an effective temperature of 3 mK, corresponding to a mode thermal occupancy of 20 phonons. 
This nearly 1000-fold increase in the mass of an optomechanical system with respect to previous experiments near the quantum ground state opens new perspectives in the exploration of the quantum/classical border.
\end{abstract}

\maketitle

{\it Introduction -} Optomechanics was born in the early 1980s when it was realized that quantum noise and radiation-pressure-induced mirror motion would eventually limit the sensitivity of large-scale gravitational-wave interferometers \cite{Caves,Reynaud}. The field experienced a huge expanse shortly after the first demonstrations of radiation-pressure cooling of a mechanical resonator \cite{Gigan, LKB3}, taking advantage of the progress in nanofabrication, and shortly experienced another breakthrough with the experimental demonstration of the quantum ground state (QGS) of a mesoscopic 15-$\mu$m mechanical drum resonator \cite{GroundStateLehnert}. Related experiments have now been performed close to the QGS with optomechanical resonators of mass ranging from 300 fg to 40 ng \cite{GroundStatePainter,GroundStateKippenberg,RegalCooling,HarrisCooling}, firmly establishing optomechanical systems as promising experimental platforms to investigate quantum effects in mesoscopic systems, complimentary to sheer condensed-matter systems \cite{Schwab,Cleland}.

A key advantage of optomechanics is radiation-pressure cooling that allows to relax the temperature constraint to reach the QGS, usually defined by the thermal phonon occupation number $n_T=k_B T/\hbar\Omega_{\rm m}\ll1$ ($\Omega_{\rm m}/2\pi$ is the resonance frequency, $T$ the environment temperature).
This allows to envision experiments at lower frequency and hence with larger mass resonators, which could lead to the observation of deviations from standard quantum mechanics. While predictions on these deviations greatly differ between theoretical models, the Planck mass $m_{\rm P}\simeq 22\,\mu{\rm g}$ is generally an interesting target regime. Foreseen experiments include tests of modified commutation relations \cite{Planck1, Marin}, quantum superpositions and non-classical states of massive objects \cite{Aspelmeyer2018,Sillanpaa2021,Aspelmeyer2021} or tests of spontaneous wavefunction collapse models \cite{Planck2,Steele}.

To date, very few experiments have been performed with truly macroscopic systems with a mass $m$ above the Planck mass, and all far from the QGS \cite{LKB1,LKB2,LKB4,Mavalvala}. 
Vibration modes of plano-convex resonators (typically, cm-scale diameter and mm-scale thickness) \cite{LKB1,LKB2,LKB4} suffer from effective masses easily three orders of magnitude above $m_\mathrm{P}$ and rather modest mechanical quality factors $Q$, mostly limited by coating and clamping losses. 
Suspended miniature mirrors \cite{Mavalvala} have similar issues, together with the extra classical noise due to the very low frequency ($\simeq$ 10 Hz) of their pendulum mode. 

\begin{figure*}
\begin{center}
\begin{tabular}{c}
\includegraphics[height=3.4cm]{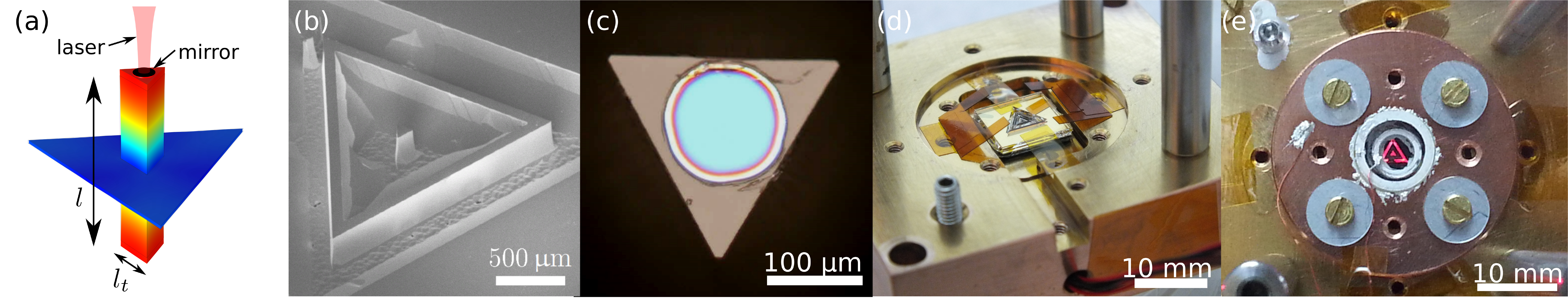}
\end{tabular}
\end{center}
\caption[example]{The mm-scale micropillar, from concept to cryogenic operation. (a) Finite-element modeling of the fundamental vibration mode. Color (from blue to red) indicates total displacement. The mirror used to reflect a laser beam is shown in black. (b) Scanning electron micrograph of an etched structure, with the additional isolation frame. (c) Optical view of a 100-$\mu$m high-reflectivity mirror coated on one end-face of the micropillar. (d) 1-cm chip (with sample) clamped inside the optical cavity mount. (e) View of the fully-assembled optomechanical cavity. The resonator is lit by a red LED.}
\label{fig:Mems}
\end{figure*}

{\it Millimeter-scale micropillar, Concept -} 
To overcome these issues, we have designed a novel resonator shaped as a micropillar\cite{APL2011, PillarPatent} as shown in Fig.~\ref{fig:Mems}(a). 
For the fundamental longitudinal mode, the displacement profile is predominantly flat across the end face, with the strain gradient along the pillar axis. 
This geometry allows for the clamping of the resonator at a longitudinal node location, and to coat the optical mirror used to probe mechanical motion on the end face, i.e. in a stress-free region. 
Thus, both clamping loss and coating loss are expected to be very low, despite the poor intrinsic mechanical quality factor of the coating layers, and we can expect a mechanical quality factor $Q$ mostly limited by the bulk material's internal losses.

A figure of merit for QGS cooling is the effective phonon occupation number $n_\mathrm{eff}\propto m \Omega_{\rm m}/Q$ \cite{MarquardtSidebandCooling} reached in a  cooling experiment, where $m$ is the effective mass of the mechanical oscillator. 
In a simple 1D model of the pillar \cite{SM} with constant $Q$, this ratio is independent of the pillar length $l$. 
The length is then chosen to set the resonance frequency of the fundamental compression mode $\Omega_{\rm m}/2\pi\simeq c_{\rm s}/2l$ ($c_{\rm s}$ is the speed of sound \cite{SM}) in the MHz range, away from technical noises such as acoustic noise, ground vibrations, and classical laser noise. 
Given the typical values of $c_{\rm s}$ in solids, $l$ is in the mm range. 
A high aspect ratio of the pillar, i.e. $l\gg l_{\rm t}$ (see Fig. \ref{fig:Mems}(a)), is advantageous to limit mechanical loss from the mirror coating \cite{SM}. 
However, for low-loss optical sensing, a transverse pillar size $l_{\rm t}$ much larger than the spot size of the laser beam (here in the 10-$\mu$m range) is needed to minimize clipping loss on the mobile mirror. We consequently pick an intermediate value $l_{\rm t}\simeq 200\,\mu$m. The effective mass $m\simeq l\times l_{\rm t}^2\times \rho$, where $\rho$ is the density of the material, is then on the order of $m\simeq 100\,\mu{\rm g}$, in the same range as $m_{\rm P}$.
To maximize the mechanical quality factor $Q$ of the resonator, monocrystalline quartz ($c_{\rm s}=6.3\times 10^3 \,{\rm m}\,{\rm s}^{-1}$, $\rho=2.7\times 10^3 \,{\rm kg}\, {\rm m}^{-3}$) is chosen since it has the lowest known intrinsic mechanical loss ($Q > 10^9$) at cryogenic temperatures \cite{Galliou}. 
To limit the surface roughness and keep mechanical surface losses on the sides of the pillar negligible, wet etching is used along the principal crystal axes: this sets the triangular cross-section of the resonator. 
The membrane connecting the center of the pillar (node location) to the surrounding structure should be as thin as possible to achieve minimal coupling to environment. 
The target membrane thickness of the order of 10 $\mu$m is compatible with the fabrication process. 
With an additional symmetric decoupling frame (see Fig.~\ref{fig:Mems}(b)), a $Q$ above $10^6$ at room temperature is expected from finite element simulations for this design.

{\it Fabrication and optical coating -}
The fabrication at ONERA starts with superpolished 1-mm thick 1.5" high-purity monocrystalline $\alpha$-quartz wafers, with the optical axis of the quartz perpendicular to the surface. 
Superpolishing provides a roughness below 2 nm rms, which is essential for optical sensing as the high-reflectivity coated layers mimic the surface roughness of the substrate.
After applying an etching mask on both sides of the wafer, 
anisotropic etching of the structure is performed in a mixture of HF and NH$_4$F at room temperature over 24 h. 
A simple characterization setup with piezoelectric actuation and Michelson interferometer sensing is used to pre-characterize the etched resonators by ringdown measurements of the $Q$ factor. 
Additional 5-minute etching steps are performed for each sample to further reduce the membrane thickness until the $Q$ exceeds $10^6$. 
The remaining mask is then removed. 

\begin{table*}
\caption{\label{tab1}Main mechanical and optical characteristics of the experiment. Temperature-dependent values are reported at cryogenic temperature. The detection efficiency takes into account the contributions $\eta_\mathrm{opt}$ from mode matching imperfections and loss due to cryostat injection optics, and $\eta_\mathrm{cav}$ from loss inside the cavity. }
\begin{ruledtabular}
\begin{tabular}{lcclcc}
\textbf{Mechanical resonator}&&&\textbf{Optical cavity}&&\\
\hline
Mechanical resonance freq.&$\Omega_{\rm m}/2\pi$&3.58 MHz&Maximum incident power&$P_{\rm max}$&25 $\mu$W\\
Mechanical quality factor&$Q$&$7\times 10^7$&Optical finesse&$\mathcal{F}$&79,000\\
Effective mass&$m$&33 $\mu$g&Cavity length&$L$&58 $\mu$m\\
Cryostat base temperature&$T$&50 mK&Cavity bandwidth (HWHM)&$\Omega_{\rm cav}/2\pi$&16.3 MHz\\
&&&Detection efficiency&$\eta_\mathrm{opt} \times\eta_\mathrm{cav}$&$0.9 \times 0.5$
\end{tabular}
\end{ruledtabular}
\end{table*}

For the optical coating, the coating area is firstly defined by transferring a dry-film photoresist with a circular hole onto the pillar.
The coating process of a dielectric Bragg mirror is carried out at Laboratoire des Mat\'eriaux Avanc\'es with a Veeco ion-beam-sputtering machine.
The coating is made of 20 doublets and the total thickness amounts to 7 $\mu$m for a residual transmission below 1 ppm. 
The dry film resist is pulled off, and the coated resonator cleaned.
A final 10-h bakeout is performed to lower the typical optical loss from 40 ppm down to 25 ppm.

{\it Room temperature properties of the coated resonator -}
A mechanical response measurement and a ringdown measurement respectively confirm the expected values of $\Omega_{\rm m}$ and $Q$ \cite{SM}.
Furthermore, the resonator mass $m$ can be inferred from the observation of the calibrated Brownian motion with a Michelson interferometer, using the equipartition theorem:
\begin{equation}\label{EquipartitionTheorem}
  \Delta x^2=\frac{\hbar}{m \Omega_\mathrm{m}}\left( n_T + \frac{1}{2}\right) \approx \frac{k_B T}{m\Omega_{\rm m}^2}\,.
\end{equation}
The exact temperature of the resonator in the presence of laser light absorption is inferred from the previously characterized temperature dependence of the mechanical resonance frequency. The measurements yield an effective mass $m = (33.5 \pm 1)\, \mu$g, in excellent agreement with both the theoretically predicted mass $m_\mathrm{th}=\rho l l_\mathrm{t}^2 \sqrt{3}/8=33\,\mu{\rm g}$ (neglecting the presence of the membrane and decoupling shield around the micropillar) and the result of COMSOL simulations for the full geometry \cite{SM}.

\begin{figure}[b]
\begin{center}
\begin{tabular}{c}
\includegraphics[width=8.5 cm]{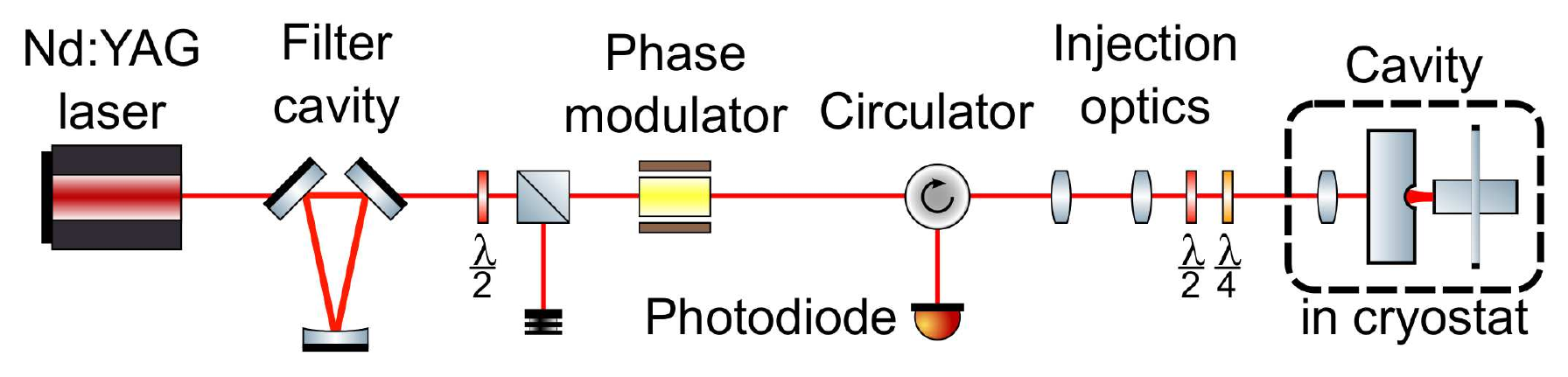}
\end{tabular}
\end{center}
\caption[example]{Experimental setup used to probe and cool the micropillar. 
A Nd:YAG laser is filtered before entering the optomechanical cavity. 
The displacement signal is extracted by direct detection of the reflected light and calibrated with a phase modulator.} \label{fig:Setup}
\end{figure}

{\it Optomechanical cavity -}
The mechanical resonator is used as the end mirror of a linear cavity (see Fig. \ref{fig:Setup}), both to probe its displacement and to reduce its effective temperature with radiation-pressure cooling \cite{LKB3, Gigan, GroundStateLehnert, GroundStatePainter,GroundStateKippenberg,HarrisCooling,RegalCooling}.
Compatibility with the mechanical resonator requires a very small optical waist (with a 6-fold margin with respect to the mirror diameter, to limit clipping losses below 1 ppm) and a bandwidth in the MHz range. 
We have opted for a compact design with a very short cavity ($L\simeq 100\,\mu{\rm m}$) and a small radius of curvature ($r_0\simeq$ 1 mm) of the coupling mirror \cite{APL2014}. 
Due to the limited tuning range of the laser source, the cavity is equipped with piezoelectric elements to allow both for tuning and locking of the cavity resonance to the laser frequency. 

The mechanical design both guarantees optical parallelism and allows for accurate translation of the resonator perpendicular to the optical axis to provide a fine centering of the resonator. 
The cavity is first fully assembled at room temperature, embedded into the open cryostat, and aligned {\it in situ} before closing and cooling down the cryostat. The cylindrical symmetry of the different cold stages of the cryostat ensures that alignment and mode-matching are not significantly altered during this process. 
We have experimentally obtained a very high finesse $\mathcal{F}= 95,000$ at room temperature, corresponding to a cavity round-trip loss of 66 ppm, with $\mathcal{T}=35$ ppm due to the coupling mirror transmission, and the remaining $\mathcal{L}=31$ ppm attributed to scattering and absorption loss of both mirrors. 
About 90\% of the light leaving the cavity arrives at the photodetector, the loss mainly occuring at the cryostat and circulator apertures. 

{\it Optomechanical sensing setup -} 
The laser beam, a Nd:YAG laser at $\lambda=1064\,\mathrm{nm}$, is sent through a triangular filtering cavity with a 30-kHz bandwidth, which suppresses its classical phase and intensity fluctuations at the 3.6-MHz frequency of interest by $\sim40$ dB, and furthermore provides spatial filtering of the beam. 
The laser source therefore delivers a $\mathrm{TEM}_{00}$ Gaussian mode with well-defined intensity and wavelength that is mode matched to the high-finesse cavity by focusing lenses. 
The laser frequency is finally locked to a given detuning $\Delta$ with respect to the cavity resonance. The resonator displacements are monitored via direct detection of the reflected laser beam by an avalanche photodiode.

{\it Cryogenic operation -} To decrease the initial thermal occupation of the mechanical oscillator, we cool its environment temperature to 50 mK in a commercial dilution refrigerator. 
A measurement of the free spectral range of the cryogenic cavity with a tunable laser reveals that thermal contraction during cool-down shortens the cavity length to 58 $\mu\mathrm{m}$. A slightly lower finesse of $\mathcal{F}=79,000$ ($\mathcal{T}=39$ ppm, $\mathcal{L}=40$ ppm, see \cite{SM}) is observed at low temperature that we attribute to contraction-related misalignment of the cavity. 
Cryogenic operation increases the resonance frequency of the mechanical oscillator by 1.4 \% due to thermal contraction and temperature-dependent elastic properties of quartz \cite{SM}. 
The strongly decreased bulk damping in mono-crystalline quartz at sub-K temperatures \cite{Galliou} has allowed us to observe $Q$-factors of $7 \times 10^7$, significantly higher than the room-temperature values but eventually limited by clamping loss \cite{SM}. 

{\it Radiation-pressure cooling -}
The thermal occupancy of the mechanical oscillator can be further reduced through radiation-pressure cooling, where the optomechanical anti-Stokes process is favored over the Stokes process by detuning the laser with respect to the cavity resonance \cite{Gigan,LKB3,GroundStateLehnert, GroundStatePainter, GroundStateKippenberg, HarrisCooling,RegalCooling}. 
As the resolved sideband regime $\Omega_\mathrm{m} \gg \Omega_\mathrm{cav}$ is impractical in our case because a long cavity would imply larger clipping loss on the coupling mirror, the system is operated in the Doppler regime \cite{RegalCooling}. 
We therefore lock the laser beam with the side-of-fringe technique on the red side of the optical resonance at a detuning of $\Delta \approx -0.77\, \Omega_\mathrm{cav}$, where efficient Doppler cooling is combined with a nearly optimal quantum backaction limit  $n_\mathrm{min} = 2.4$
for the minimum thermal occupation number.

\begin{figure}[t]
\begin{center}
\begin{tabular}{c}
\includegraphics[width=8.5 cm]{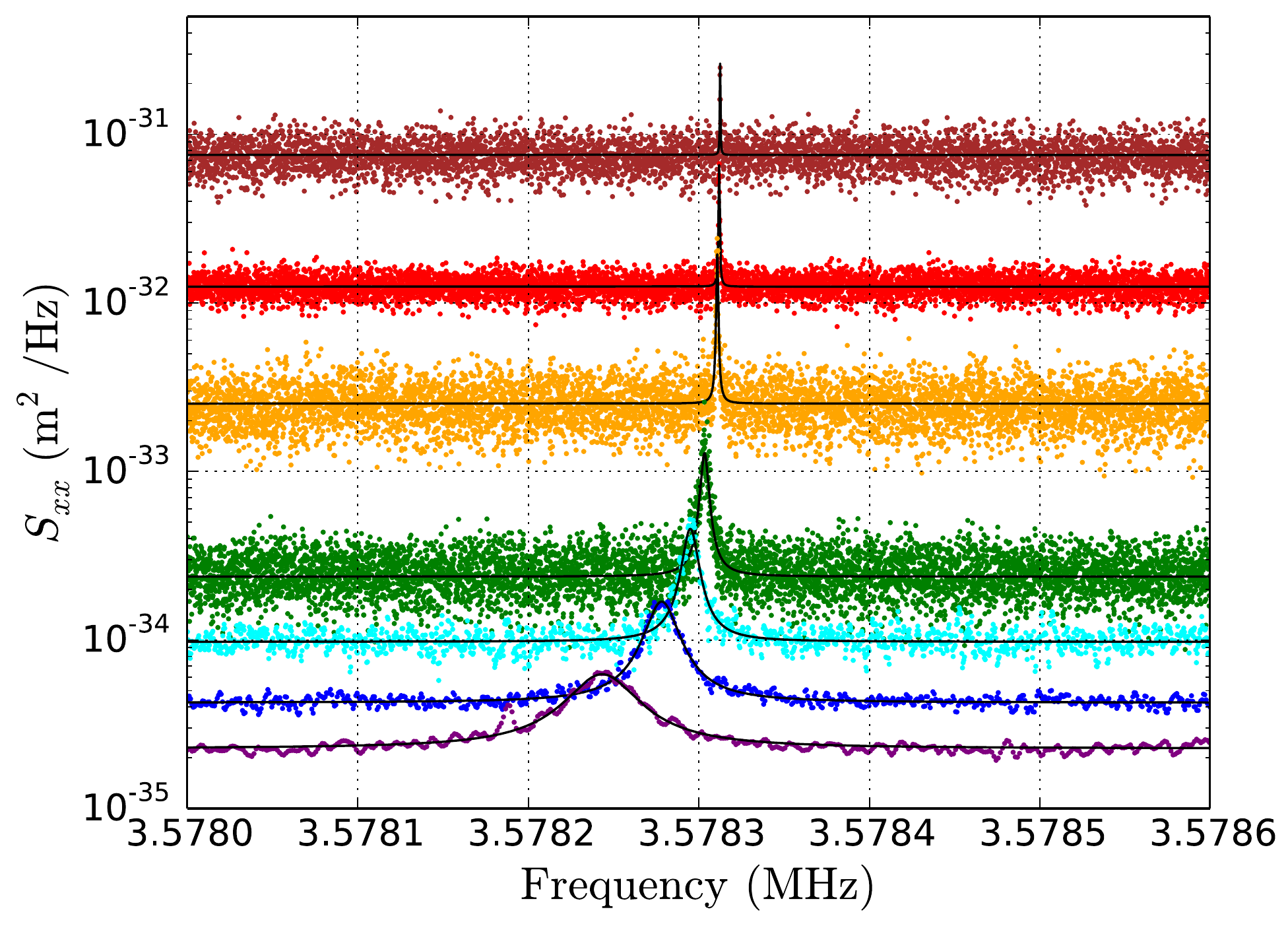}
\end{tabular}
\end{center}
\caption[example]{Noise spectra of the micropillar displacements for a detuned cavity. 
The different colored curves correspond (from top to bottom) to an incident power of 0.2, 0.4, 0.8, 3.2, 6.4, 12.5 and 25.0 $\mu$W. 
The mechanical resonance peak becomes both wider and lower for increasing power as expected. 
The effective temperature and resonance width can be extracted from Lorentzian fits (black lines). The bottom curve corresponds to an average occupation number of 20 phonons.} \label{fig:Spectra}
\end{figure}

\begin{figure}[t]
\begin{center}
\includegraphics[width=8.5 cm]{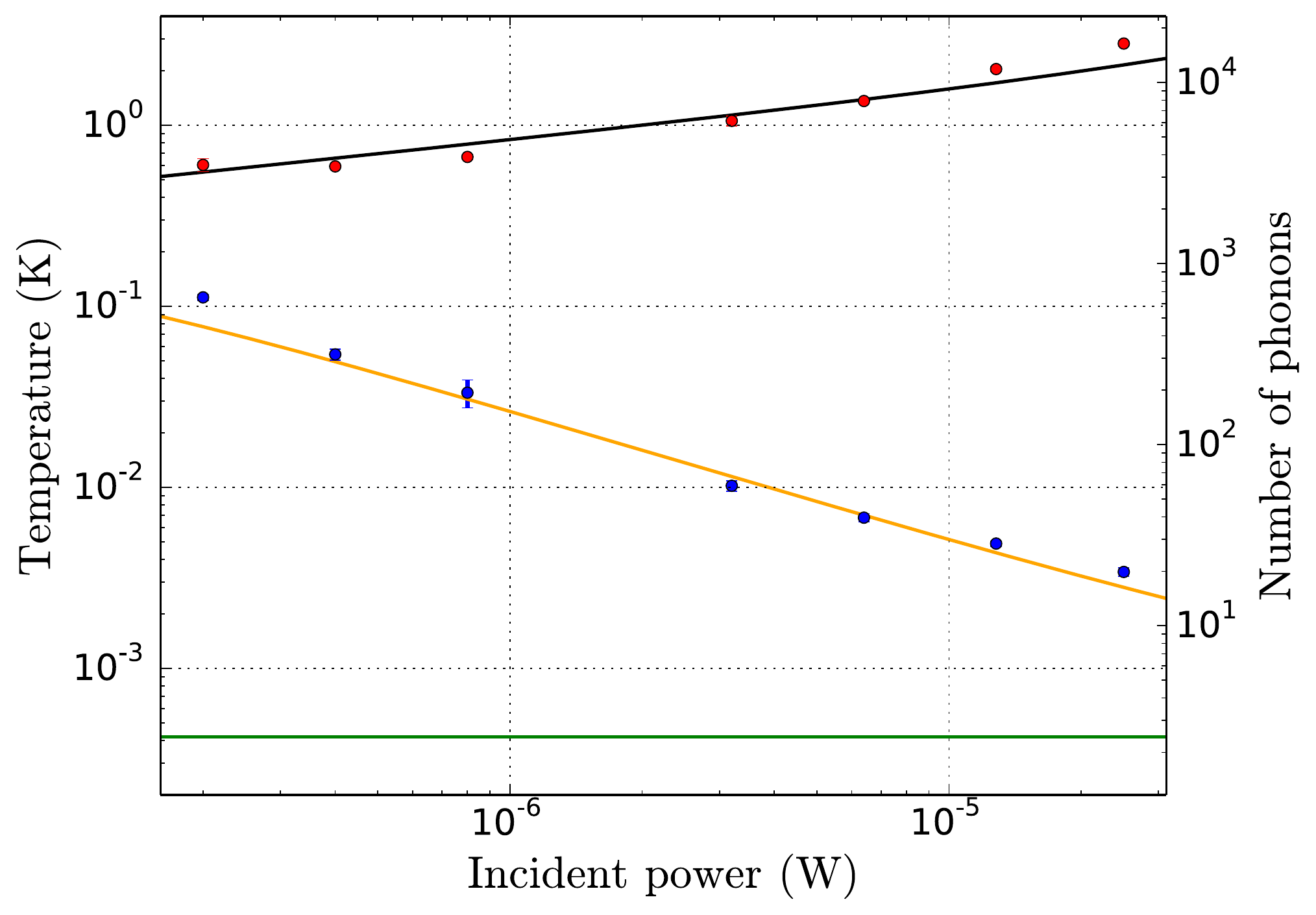}
\end{center}
\caption[example]{Measured effective mode temperature $T_\mathrm{eff}$ (blue dots, left scale), effective phonon number $n_\mathrm{eff}$ (blue dots, right scale), and deduced environment temperature $T_\mathrm{env}$ (red dots, left scale) as a function of injected laser power, with corresponding models (orange and black lines, see text). 
The green line represents the quantum backaction limit. 
Error bars indicate variations among successive measurements. 
} \label{fig:Results}
\end{figure}

Noise spectra of the reflected photocurrent as a function of the injected laser power are acquired via direct detection and calibrated with a phase modulation at 3.6 MHz on the incident laser beam. The resulting spectra are shown in Fig. \ref{fig:Spectra}. 
For each injected power, we also perform ringdown and optomechanically induced transparency (OMIT) response measurements  \cite{OMIT}. Lorentzian fits of the spectra, exponential fits of the ringdown traces, and fits of the OMIT measurements all yield effective mechanical damping rates in mutual agreement and linear as a function of the injected power \cite{SM}. 
At the maximum employed power of $25\,\mu\mathrm{W}$, we achieve an 800-fold increase of the intrinsic mechanical damping rate. 
The injection of higher optical power into the cavity is prevented by the onset of an optomechanical instability related to low-frequency mechanical modes of the sample. The noise floor of the spectra is dominated by the darknoise of the photodiode for all powers. 
Fig. \ref{fig:Results} shows the effective temperatures $T_\mathrm{eff}$ and thermal occupation numbers $n_\mathrm{eff}$ obtained with Eq. (\ref{EquipartitionTheorem}) from Lorentzian fits of the spectra. 
The lowest value obtained for the highest injected power is $n_\mathrm{eff} = 20$. 

{\it Absorption heating -}
To estimate the heating of the micropillar due to laser light absorption, we deduce the environment temperature $T_\mathrm{env} = T_\mathrm{eff} \gamma_\mathrm{eff} / \gamma_\mathrm{m}$ by assuming a power-independent intrinsic mechanical linewidth $\gamma_\mathrm{m} = \Omega_\mathrm{m}/Q$ and accounting for the expected amount of laser-cooling $\gamma_\mathrm{opt}$ with the observed effective linewidth $\gamma_\mathrm{eff} = \gamma_\mathrm{m} + \gamma_\mathrm{opt}$ extracted for each measured spectrum. 
To model the observed power-dependence, we first estimate the boundary scattering-limited heat conductivity through the quartz membrane that supports the micropillar from a Debye model \cite{SM}. Literature values for the temperature-dependent heat capacity of quartz \cite{HeatCapacity} and a phonon mean free path of $\Lambda = 24\,\mu$m deduced from the geometry of the membrane yield an estimated thermal conductivity of $\kappa = 8 \times 10^{-2}\,\mathrm{W}\,\mathrm{m}^{-1}\,\mathrm{K}^{-1}$ at 1 K, with a $T^3$-dependence. This results in a $P^{1/4}$-dependence of the pillar temperature on the dissipated power $P$, both in a purely analytical model and in a finite-element simulation which mutually agree within 10 \% \cite{SM}. The only free parameter of the model, the fraction of absorbed light power dissipated on the micropillar, is adjusted to 20 \% to obtain a good fit between measurements and model (red dots and black line in Fig. \ref{fig:Results}). The modeled environment temperature can then be used to accurately predict the observed effective temperature by assuming standard radiation-pressure cooling without any additional adjustable parameter (blue dots and orange line in Fig. \ref{fig:Results}). The slight underestimation of the force noise at the highest powers by our model is a result of the temperature-dependence of the intrinsic quality factor. A complementary measurement at 4 K with zero detuning gives a values of $Q=2.5\times10^7$, and is consistent with the observed deviation at high power in Fig. \ref{fig:Results}.
At the highest power, the thermal occupation number is about one order of magnitude above the quantum backaction limit (green line in Fig. \ref{fig:Results}). 

{\it Conclusion} -
We have presented an optomechanical system with a mass nearly 1,000 times above the heaviest system previously demonstrated in the quantum ground state, cooled to an effective thermal occupation number of 20 phonons. Further cooling is limited by heating of the sample by absorbed laser light. An analytical model for the heat transport and optomechanical effects with only one adjustable parameter successfully explains the observed effective temperature. Future work will aim at a further reduction of the thermal motion of the oscillator through feedback cooling \cite{LKB2,FeedbackKippenberg,FeedbackSchliesser}, and the demonstration of complementary experimental signatures of quantum motion \cite{PainterAsymmetry} of our macroscopic optomechanical system. While creating a large-scale cat state of such a resonator is still a long way ahead, we note the micropillar is one of the very few mechanical systems where gravitational decoherence could be demonstrated according to the model developed in \cite{Steele}.

\smallskip
We acknowledge A. G. Kuhn for his work on an earlier version of this experiment. 
Fabrication has been partially carried out at Universit\'e de Paris with support from Pascal Filloux, Christophe Manquest, and St\'ephan Suffit. 
We thank an anonymous referee for constructive remarks. 
This research has been funded by the Marie Curie Initial Training Network Cavity Quantum
Optomechanics and ANR program ANR-15-CE30-0014 ExSqueez. 
S. C. acknowledges support from the European Commission through Marie Curie Fellowship IIF Project SQZOMS No. 660941.


\end{document}


\title{Supplementary Information for \\Laser cooling of a Planck mass object close to the quantum ground state}

\author{L. Neuhaus}
\author{R. Metzdorff}
\author{S. Zerkani}
\author{S. Chua}
\author{J. Teissier\footnote{Current Affiliation: II-VI Laser Enterprise, Binstrasse 17, CH-8045 Zurich, Switzerland}}
\author{D. Garcia-Sanchez\footnote{Current affiliation: Institut des NanoSciences de Paris, Sorbonne Universit\'e, CNRS-UMR 7588,  F-75005, Paris, France}}
\author{S. Del\'eglise}
\author{T. Jacqmin}
\author{T. Briant}
\affiliation{Laboratoire Kastler Brossel, Sorbonne Universit\'e, ENS - Universit\'e PSL, Coll\`ege de France, CNRS, Paris,
France}
\author{J. Degallaix}

\author{V. Dolique\footnote{Current Affiliation: Univ Lyon, ENS de Lyon, Univ Claude Bernard Lyon 1, CNRS, Laboratoire de Physique, F-69342 Lyon, \author{C. Michel}
\author{L. Pinard}\affiliation{Universit\'e de Lyon, Universit\'e Claude Bernard Lyon 1, CNRS, Laboratoire des Mat\'eriaux Avanc\'es (LMA), IP2I Lyon / IN2P3, UMR 5822, F-69622 Villeurbanne, France}
France}}
\author{G. Cagnoli}
\affiliation{Universit\'e de Lyon, Universit\'e Claude Bernard Lyon 1, CNRS, Institut Lumi\`ere
Mati\`ere, F-69622, Villeurbanne, France}
\author{O. Le Traon}
\author{C. Chartier}
\affiliation{ONERA—The French Aerospace Lab, F-91123 Palaiseau Cedex, France}
\author{A. Heidmann}
\author{P.-F. Cohadon}
\email{cohadon@lkb.upmc.fr}
\affiliation{Laboratoire Kastler Brossel, Sorbonne Universit\'e, ENS - Universit\'e PSL, Coll\`ege de France, CNRS, Paris,
France}

\maketitle

In this Supplementary Information, we provide more details regarding four key aspects of our experiment. The first section deals with the modeling and characterization of the mechanical resonator. The second section lines out the calculation of the temperature-dependent thermal conductivity and the resulting model for the heating by light absorbed by the resonator. The third section presents the characterization of the power-dependent optomechanical damping rate that allows to more accurately model optomechanical effects in our experiment with an experimental estimate of the transmission of the input mirror of our cavity. The last section presents measurements of the laser classical noise and allows us to conclude that this noise has no measurable effect on our experiment.

\section{Mechanical resonator}
\begin{table}
\begin{ruledtabular}
\begin{tabular}{lccc}
&$\Omega_{\rm m}/2\pi$ (MHz)&$m$ ($\mu$g)& $Q$ \\
\hline
Theory & 3.15 &  33.0 & - \\
Simulation&3.62& 30.0 & $1.2\times10^8$\\
Measurement& 3.53 & 33.5 & $7\times 10^7$ \\
\end{tabular}
\end{ruledtabular}
\caption{\label{tab1}Mechanical resonance frequency $\Omega_\mathrm{m}/2\pi$, effective mass $m$, and mechanical quality factor $Q$ from analytical estimates, finite-element simulation, and measurement at room temperature. The simulated $Q$ only takes into account clamping loss. The measured $Q$ is therefore reported for cryogenic temperature where all other loss channels are minimum as to obtain an upper bound for the achieved clamping loss. The different methods are in good mutual agreement. }
\end{table}
\begin{figure}
\begin{center}
\begin{tabular}{c}
\includegraphics[width=8.5 cm]{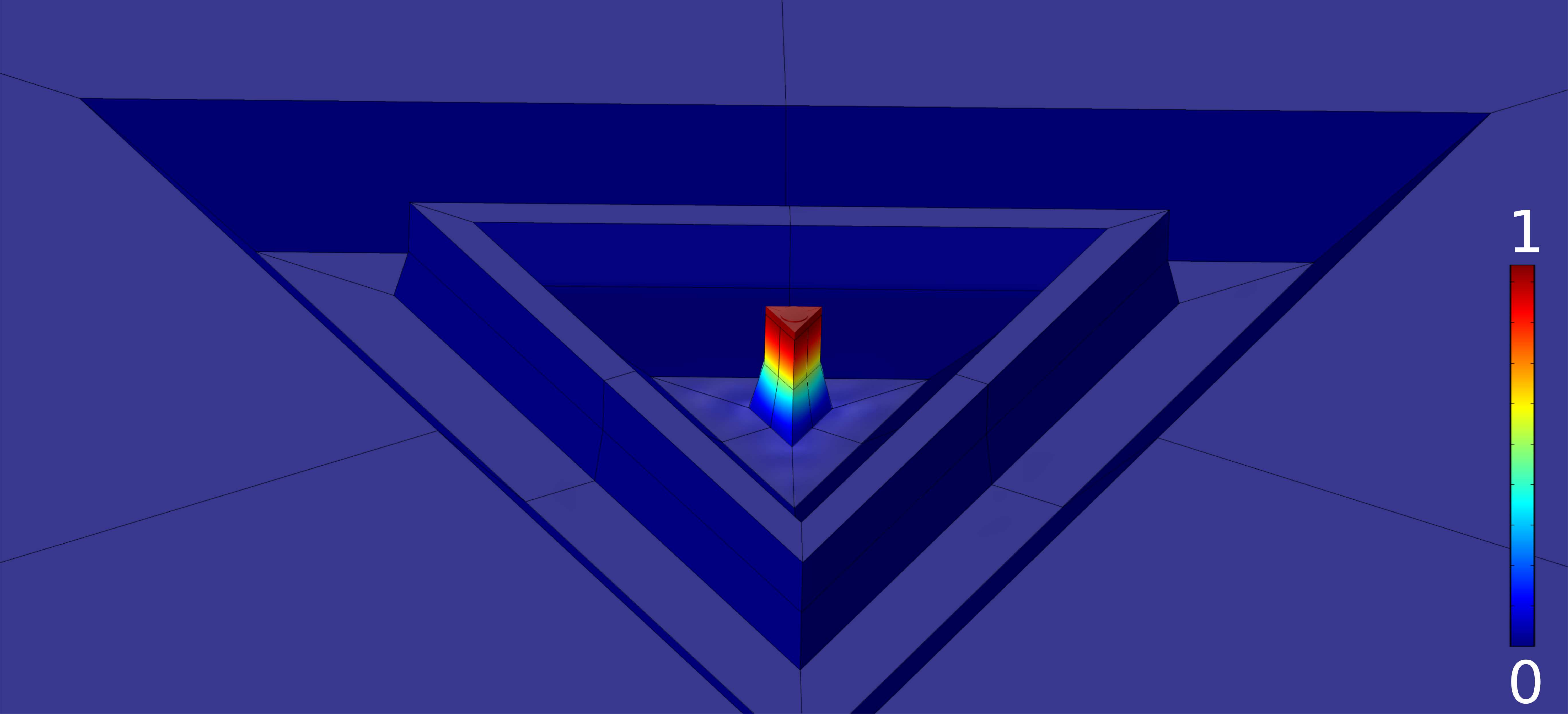}
\end{tabular}
\end{center}
\caption[example]{COMSOL simulation of the fundamental compression mode of the micropillar resonator. Color encodes total displacement in arbitrary units. The outermost frame is defined as perfectly matched layer for the simulation of clamping loss. } 
\label{Fig:fundamentalmode}
\end{figure}
\begin{figure}
\begin{center}
\begin{tabular}{c}
\includegraphics[width=8.5 cm]{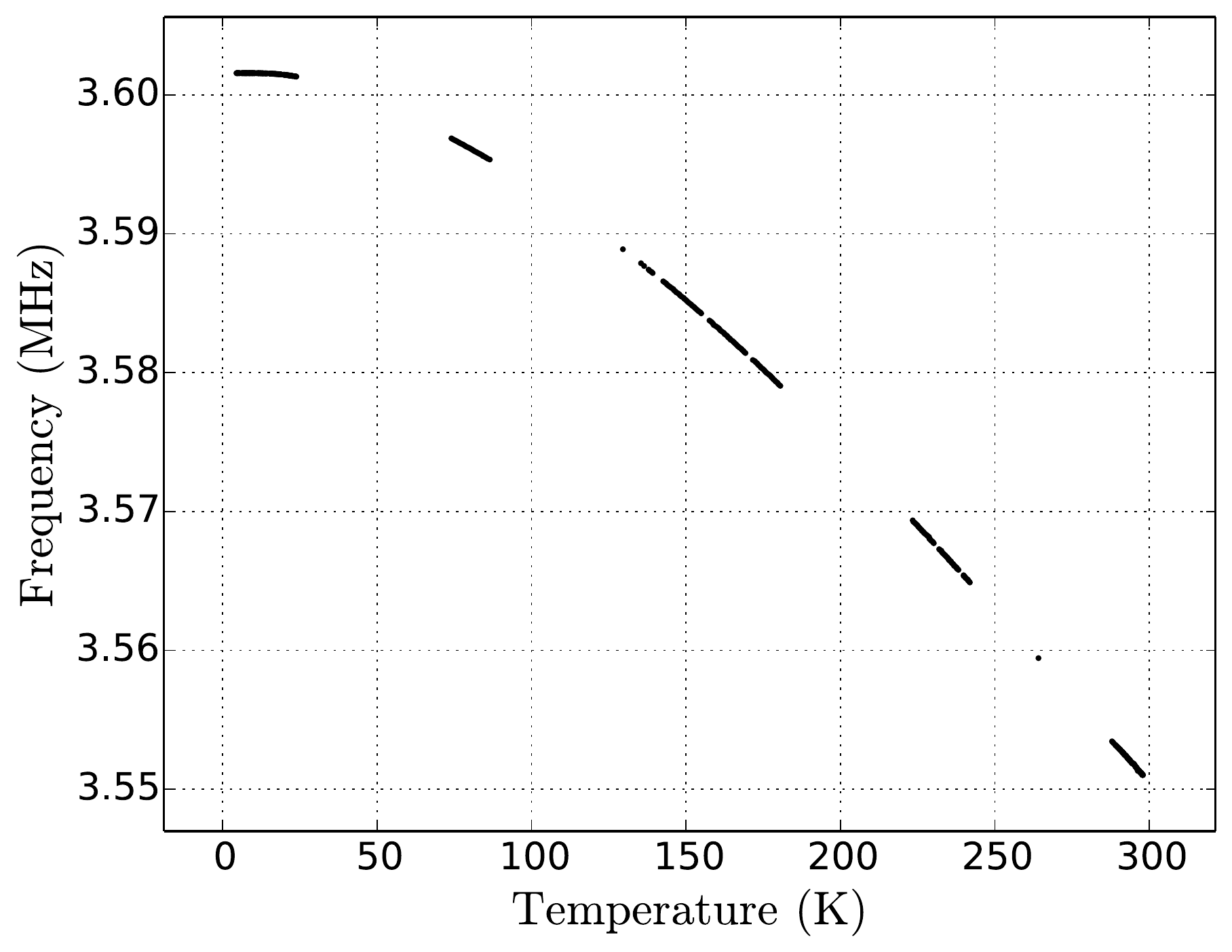}
\end{tabular}
\end{center}
\caption[example]{Measured resonance frequency versus cryostat temperature for one micropillar sample while the cryostat is heated up. At room temperature, the slope is -248 Hz/K. } 
\label{Fig:f_vs_T}
\end{figure}
\begin{figure*}
\begin{center}
\begin{tabular}{c}
\includegraphics[width=17cm]{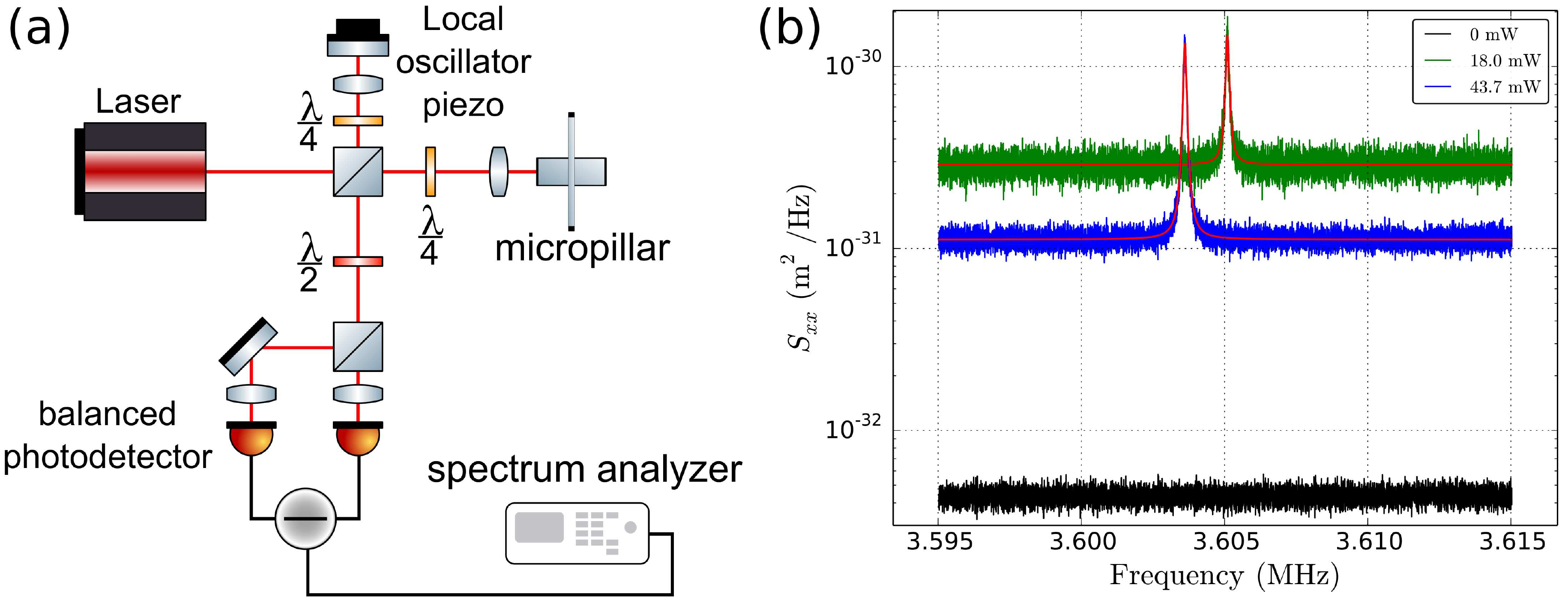}
\end{tabular}
\end{center}
\caption[example]{
(a) Schematic of the Michelson interferometer used to determine the effective mass of the mechanical resonator. 
The balanced photodetector is a home-built electronic circuit that directly subtracts the photocurrents of two high quantum efficiency InGaAs photodiodes before amplification, which rejects classical laser noise by more than 60 dB in this measurement. 
The low-frequency part of the photocurrent is further amplified and fed to the local oscillator piezo to stabilize the differential interferometer path length to achieve maximum sensitivity for the pillar displacement. 
The electronic spectra of the differential photocurrent are calibrated from the interference fringe amplitude when a voltage ramp is applied to the local oscillator piezo. 
(b) Measured thermal noise spectra of the mechanical oscillator at room temperature and atmospheric pressure with different optical powers injected into the interferometer (green and blue traces). 
Red lines are Lorentzian fits of the spectra. The black dark noise trace was converted to displacement units with the conversion factor of the blue trace. } \label{Fig:massmeasurement}
\end{figure*}
\begin{figure*}
\begin{center}
\begin{tabular}{c}
\includegraphics[width=17.0cm]{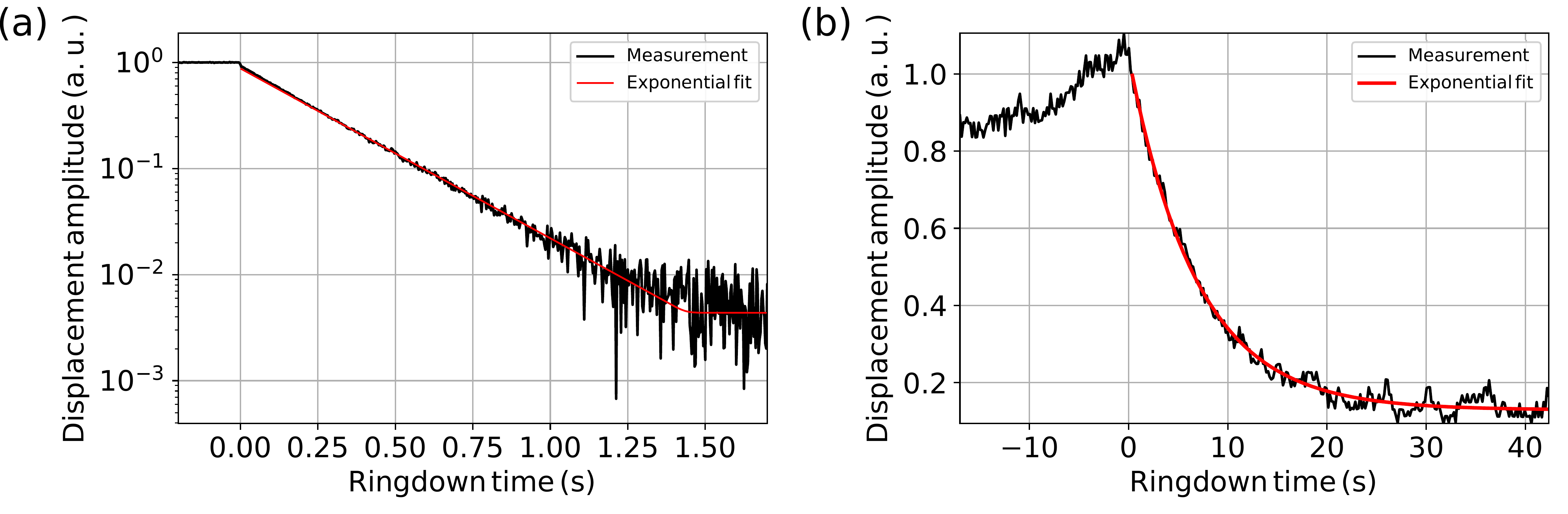}
\end{tabular}
\end{center}
\caption{(a) Ringdown measurement at room temperature. The micropillar is mounted on a piezoelectric actuator and driven to a large vibration amplidude. After disabling the drive signal at the time $t=0\,\mathrm{s}$, the decay of this motion is monitored by the interferometer setup shown in Fig. \ref{Fig:massmeasurement} (a). An exponential fit with a constant offset to account for the measurement noise floor yields a mechanical quality factor of $Q=3.0\times 10^6$. 
(b) Ringdown measurement at a cryogenic temperature $T=50\,\mathrm{mK}$. See text for details. The exponential fit of the data yields a mechanical quality factor of $Q=7.6\times10^7$.}
\label{Fig:ringdownplots}
\end{figure*}
\begin{figure*}
\begin{center}
\begin{tabular}{c}
\includegraphics[width=17.0cm]{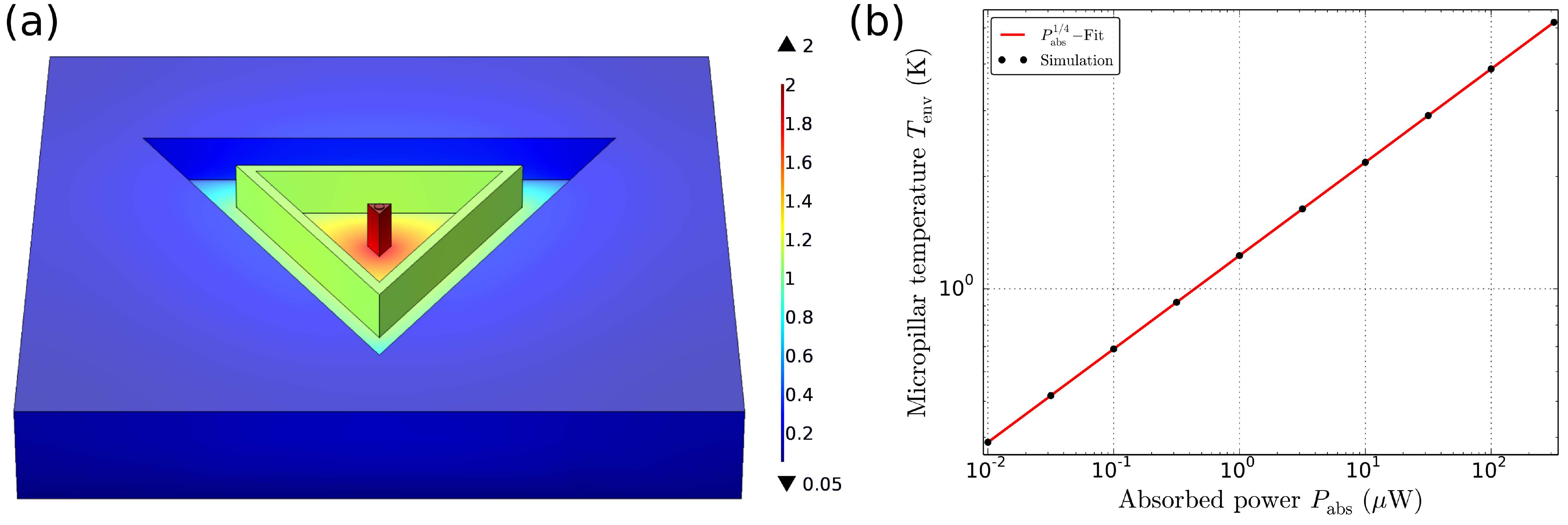}
\end{tabular}
\end{center}
\caption{(a) Simulated temperature field for a membrane thickness of 10 $\mu$m and a power of 5 $\mu$W absorbed by the round mirror on top of the micropillar. The color bar indicates the temperature in K. 
(b) Simulated mean temperature of the central micropillar for various absorbed powers (black dots). The theoretically expected power law yields an excellent fit of the simulated temperatures (red line). }
\label{fig:temperaturesimulation}
\end{figure*}
\begin{figure*}[t]
\begin{center}
\begin{tabular}{c}
\includegraphics[width=17cm]{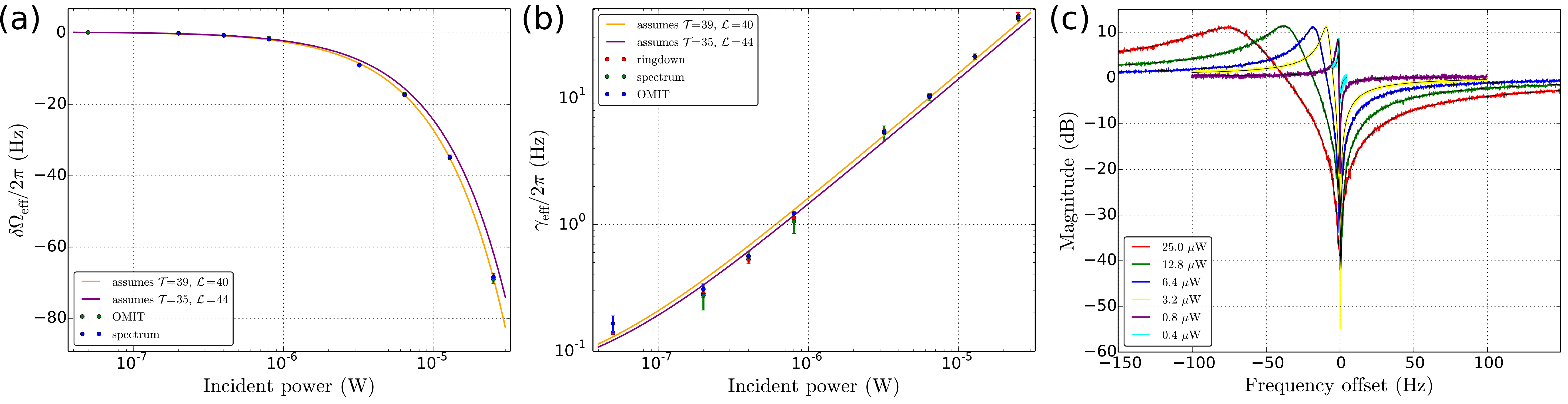} 
\end{tabular}
\end{center}
\caption[example]{(a) Optical spring $\delta \Omega_\mathrm{eff}$ and (b) effective damping $\gamma_\mathrm{eff}$ versus incident optical power obtained from different measurements (colored dots). The measurements agree well with linear models, assuming different transmissions for the input coupling mirror (yellow and purple lines, see main text). (c) Magnitude of the measured transfer function across the intrinsic mechanical resonance frequency from an phase modulation on the injected beam to the reflected power from the optomechanical cavity for a red-detuned laser and different powers. Curves are normalized to 0 dB at frequencies far from the mechanical resonance. Thin black lines are complex Lorentzian (Fano lineshape) fits of the measurements. The sharp absorption dip at the center frequency is characteristic of optomechanically induced transparency. } \label{Fig:optdamping}
\end{figure*}
\begin{figure*}
\begin{center}
\begin{tabular}{c}
\includegraphics[width=17.0cm]{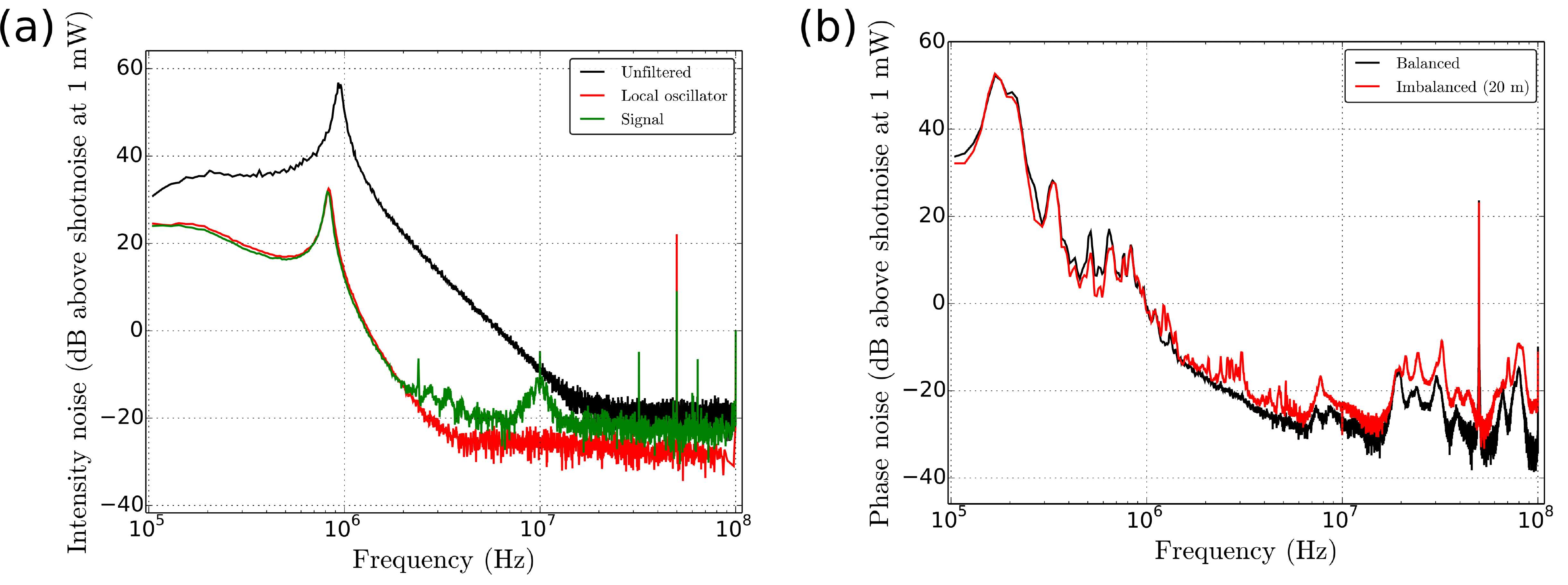}
\end{tabular}
\end{center}
\caption{(a) Intensity noise of the Nd:YAG laser source before and after the 60-kHz-linewidth filter cavity employed for all measurements. The filtered laser noise was measured in the local oscillator (LO) path of the setup. The beam that is injected into the optomechanical cavity (signal) has a slightly higher intensity noise due to an amplitude modulator present in the signal path that produces additional noise. 
(b) Phase noise of the Nd:YAG laser source after the filter cavity measured by locking the interference signal between the local oscillator and signal paths at the mid-fringe position. To estimate the frequency noise of the laser source with an imbalanced Mach-Zehnder interferometer, the measurement is repeated with a 20-m single-mode fiber delay line in one interferometer arm. All traces are normalized to the quantum shot noise level that is measured with a total power of 1 mW incident on the balanced photodetector. }
\label{Fig:lasernoise}
\end{figure*}
The main properties of our mechanical oscillator obtained from analytical considerations, finite-element simulation, and characterization measurements are shown in Table \ref{tab1}. 
Fig. \ref{Fig:fundamentalmode} shows the geometry used for the finite-element simulation and the resulting displacement field for the fundamental compression mode of the micropillar. 

As mentioned in the article, the mechanical frequency $\Omega_\mathrm{m}/2\pi \approx c_\mathrm{s}/2l$ is expected to be proportional to the ratio between the speed of sound in the micropillar $c_\mathrm{s}=\sqrt{E/\rho}=6300\,\mathrm{m}\,\mathrm{s}^{-1}$ and the resonator thickness $l = 1\,\mathrm{mm}$, where $E$ denotes the elastic modulus of $\alpha$-quartz in the direction of the optical crystal axis and $\rho=2650\,\mathrm{kg}\,\mathrm{m}^{-3}$ the density. The remaining difference ($<15\,\%$) between this theoretical model and the finite-element simulation and measurement arises from not taking into account the finite transverse size of the micropillar in the analytical model, which would increase the resonance frequency closer to the simulated and measured values. The 3\%-disagreement between measured and simulated frequency is mainly attributed to the difference between simulated and actual sample thickness. When the resonator is cooled from room temperature to 4 K, the resonance frequency increases by about 50 kHz (Fig. \ref{Fig:f_vs_T}).

The effective mass $m$ of a mechanical mode in an optomechanical system is defined as the mass of a mirror on a spring that would yield the same optomechanical response as the mechanical mode \cite{HeidmannMass}. 
A simple analytical estimate of the effective mass of the micropillar's compression mode is obtained by approximating the mode as the fundamental thickness oscillation of a prism with equilateral triangular crosssection. This results in an effective mode mass $m = m_\mathrm{prism}/2 = 33 \,\mu\mathrm{g}$ with the physical prism mass $m_\mathrm{prism} = \rho l l_\mathrm{t}^2 \sqrt{3}/4$
 and triangle side length $l_\mathrm{t}=240\,\mu\mathrm{m}$. 
The simulated displacement field in Fig. \ref{Fig:fundamentalmode} shows that this is a good approximation of the expected mode shape, which explains the good agreement between theory and simulation in Table \ref{tab1}. 

The interferometric measurement of the effective mass compares the measured Brownian motion of the oscillator (Fig. \ref{Fig:massmeasurement}) to the expectation from the equipartition theorem with a known temperature. 
Fig. \ref{Fig:massmeasurement}(b) shows that the observed resonance frequency changes when the optical power reflected by the micropillar is varied. 
With the observed frequency shift and the measured frequency dependence with temperature shown in Fig. \ref{Fig:f_vs_T}, we estimate laser heating to account for an error below 3 \% in the mass measurement. 

Fig. \ref{Fig:ringdownplots} (a) shows a  mechanical  ringdown measurement at room temperature. 
The room-temperature measurement was performed with an interferometric readout, which allows for a relatively large dynamic range of about 50 dB. 
The measured room-temperature quality factor of $3\times10^6$ is close to the limit $Q<4\times10^6$ expected for bulk crystalline quartz \cite{Warner} and therefore indicates that the mechanical loss of our samples is dominated by internal damping at this temperature. 
A measurement with cavity optomechanical readout yields similar values if cooling or heating by dynamical backaction is avoided by employing low laser powers below $1\,\mu\mathrm{W}$ and locking the laser at the cavity resonance frequency. 

At cryogenic temperature ($T<1$ K), the intrinsic quality factor of $\alpha$-quartz is expected above $10^9$ \cite{Galliou}, which makes the measured $Q$ sensitive to clamping loss variations due to different mounting conditions in our experiment. 
We therefore perform the low-temperature ringdown measurements with the sample mounted in the optomechanical cavity configuration and during the same cooling cycle as subsequent optomechanical measurements. 
Standard cavity optomechanical readout techniques such as side-of-fringe or Pound-Drever-Hall have yielded unreliable ringdown measurement results in this quality factor range since even small detunings between the laser and the optical resonance at the lowest practical laser power levels cause significant quality factor variations through dynamical backaction. 
We therefore employ a novel readout scheme where the broadening of the optical resonance by mechanical dispacements with amplitudes larger than the optical linewidth is monitored during the ringdown by periodically sweeping over the optical resonance, which keeps dynamical backaction effects at a minimum through low intracavity power and near-zero detunings averaged over each sweep cycle \cite{Neuhaus}. 
Fig. \ref{Fig:ringdownplots} (b) shows one such measurement result that yields a cryogenic quality factor of $7.5\times10^7$. 
Since the noise floor of our detection scheme is given by the integrated displacement noise of the cavity and the maximum mechanical drive amplitude is limited by the onset of nonlinear effects, the dynamic range of this measurement is smaller than in the standard interferometric ringdown measurement at room temperature. 
While cryogenic measurements during a single cooldown run are consistent, we have observed a variation among different cryostat runs with $Q$-factors ranging from $2.5\times10^7$ to $7.5\times10^7$. 
All measured quality factors at cryogenic temperature are significantly below the intrinsic quality factor of $\alpha$-quartz ($Q_\mathrm{intrinsic} > 10^9$, see Ref. \cite{Galliou}). 

To identify the dominant cryogenic loss mechanism, we consider mechanical dissipation by the dielectric mirror coating, by surrounding air, and clamping loss from surrounding support structure.  
The intrinsic quality of the dielectric mirror coating is expected in the range $10^3$ to $10^4$ \cite{Crooks}. Its impact on the overall $Q$-factor is however diluted by the small fraction of elastic energy stored within the mirror, which can be estimated by finite-element simulations. 
This yields an estimated limit of $Q=3\times10^8$ for loss from the mirror coating. 
Gas damping estimated from a pressure-dependent quality factor measurement series yields a limit of $Q=10^{12}$ for an overestimated cryogenic pressure of $p<10^{-6}\,\mathrm{mbar}$. 
Using a finite-element model, we obtain an estimate of $Q=1.2 \times 10^8$ for the clamping loss limit by surrounding the mechanical resonator structure as shown in Fig. \ref{Fig:fundamentalmode} with a perfectly matched layer that absorbs all incident acoustic radiation from the central oscillator. 
A small variation of the simulated geometry, for example a thickess variation of the central membrane or the pillar thickness, causes a significant variation of the obtained clamping loss, such that a range of plausible geometries yields the measured quality factors. 
The observed dependence on mounting conditions and cooldown run could thus be explained by a variation of mechanical stress of the substrate and the resulting clamping loss variation. 
This strong dependence jeopardizes our ability to compute the precise quality factor value of our samples, given the the imperfect knowledge of the exact geometry of our mechanical resonator, especially of the thickness and topology of the membrane supporting the central pillar, and an unknown stress distribution of the sample chip. 
Since the simulated $Q$-factor of all slightly varied geometries covers the range of measured values, we conclude that clamping loss is the limiting factor for the mechanical $Q$-factor of our samples at cryogenic temperatures. 

\section{Absorption heating model}
\begin{table}
\begin{ruledtabular}
\begin{tabular}{ c c c c }
& $\kappa$ & $C_v$  & Validity range\\
& (W m$^{-1}$ K$^{-1}$)  & (W s kg$^{-1}$ K$^{-1}$) & K\\
	\hline
Ref. \cite{Zeller1971} & $100 \left(T/1\,\mathrm{K}\right)^3$ & $5.7\times10^{-4} \left(T/1\,\mathrm{K}\right)^3$ & $0.2 - 10$ \\
Ref. \cite{Gardner1981} &$15 \left(T/1\,\mathrm{K}\right)^3$ & $5.6 \times 10^{-4}\left(T\right/1\,\mathrm{K})^3$ &  $0.1 - 3$ \\
\end{tabular}
\end{ruledtabular}
\caption{\small Literature values for the thermal conductivity $\kappa$ along the optical axis and the specific heat capacity $C_v$ of $\alpha$-quartz at low temperature. The thermal conductivity in Ref. \cite{Gardner1981} was limited by boundary scattering at the sample edges. } 
\label{quartzconstants}
\end{table}
Literature values of the thermal properties of $\alpha$-quartz at low temperature are listed in Table \ref{quartzconstants}. 
While the measurements of the specific heat capacity are in good agreement, the reported thermal conductivities vary significantly. 
This variation is a result from heat transport in $\alpha$-quartz at low temperature being ballistic, i.e. the phonons that transport heat across the bulk material are predominantly scattered by the sample boundaries instead of being scattered by lattice imperfections or other phonons. 
The thermal conductivity must therefore be estimated from the sample geometry. 
From the Debye model \cite{Ashcroft1976}, we expect the thermal conductivity $\kappa$ to be 
\begin{equation}
\kappa = \frac{1}{3} \rho C_v c_\mathrm{s} \Lambda\, ,
\end{equation}
where $C_v$ is the specific heat capacity, $c_\mathrm{s}$ the sound velocity and $\Lambda$ the phonon mean free path. 
We estimate the mean free path from the dimensions of our sample with the assumption that the thermal conductivity is limited by the membrane between the central micropillar and the sourrounding outer frame. 
The distance of a line from a point at the center of the membrane of thickness $t\approx 10\,\mu$m to the upper or lower membrane surface at an angle $\phi$ with respect to the plane is $l_\mathrm{geo}(\phi) = t/(2 \sin \phi)$. 
As $l_\mathrm{geo}$ diverges for angles near zero, we assume that the free path length is furthermore limited by the lateral sample size $l_\mathrm{max} = 10 $ mm, such that $l_\mathrm{total}^{-1} = l_\mathrm{geo}^{-1} + l_\mathrm{max}^{-1}$. 
By averaging $l_\mathrm{total}(\phi)$ over all possible angles, we numerically find for the mean free path
\begin{equation}
\Lambda = \frac{1}{\pi}\int_{-\pi/2}^{\pi/2} \frac{1}{1/l_\mathrm{max} + 2 \sin(\phi)/t}\mathrm{d}\phi \approx 24\,\mu\mathrm{m}\, .
\label{meanfreepath}
\end{equation}
With the numerical values introduced above and in Table \ref{quartzconstants}, we find a thermal conductivity 
\begin{equation}
\kappa \approx 8\times 10^{-2} \left(\frac{T}{1 \mathrm{K}}\right)^3\,(\mathrm{W}\,\mathrm{m}^{-1}\,\mathrm{K}^{-1})
\label{thconductivity}
\end{equation}
which is approximately proportional to the membrane thickness $t$ in this parameter region. 

We proceed to estimate the sample temperature assuming that the thermal resistance is dominated by heat conduction through the membrane. A simple analytical model can be derived by approximating the membrane as a ring of thickness $t$, inner radius $r_\mathrm{micropillar} \approx 115\, \mu$m and outer radius $r_\mathrm{frame} \approx 1.2\,\mathrm{mm}$. Here, the inner and outer radii are chosen such that the circumference of the rings coincides with the corresponding circumference in the triangular sample structure. The thermal gradient $\mathrm{d}T/\mathrm{d}r$ at the  radial distance $r$ from the center of the disc is related to the dissipated power $P$ that is conducted from the pillar edge at $r_\mathrm{micropillar}$ to the frame at $r_\mathrm{frame}$ as
\begin{equation}
P = - 2 \pi r t \kappa \frac{\mathrm{d}T}{\mathrm{d}r}\, .
\end{equation}
After inserting the temperature-dependent thermal conductivity from Eq. (\ref{thconductivity}), the equation can be solved by integrating from the inner to the outer disc radius over a term proportional to $r^{-1} \mathrm{d}r$ on one side and over a term proportional to $T^3\mathrm{d}T$ on the other side. We eventually obtain
\begin{equation}
\frac{T_\mathrm{micropillar}}{1\,\mathrm{K}} = \left(\frac{P}{0.4\,\mu\mathrm{W}} + \left(\frac{T_\mathrm{frame}}{1 \mathrm{K}}\right)^4\right)^{1/4} \, .
\label{analyticalheating}
\end{equation}
Since the low cryostat temperature $T_\mathrm{frame}\approx 50\,\mathrm{mK}$ only has a negligible contribution compared to the heating term in our parameter regime, the micropillar temperature $T_\mathrm{micropillar}$ should scale with the fourth root of the absorbed power and the inverse square root of membrane thickness. 
In addition to the analytical model, we have performed a finite-element simulation of the temperature field at various powers with the same material parameters (Fig. \ref{fig:temperaturesimulation}). A power-law fit of the simulation results yields the expected $P^{1/4}$-dependence with a pre-factor within 10\% to the one derived from the analytical model. 
We expect the power $P$ dissipated on the micropillar to be proportional to the optical power absorbed by the optomechanical cavity. As discussed in the main text, the experimentally obtained micropillar temperatures $T_\mathrm{env}$ are consistent with our model when it is assumed that 20 \% of the optical power is dissipated by the micropillar. 

Marginally better agreement between measurement and theory can be achieved by modeling the micropillar temperature with an arbitrary power law, which yields a dependence as $T_\mathrm{env} \propto P^{2/3} + const$. The exponent being larger than the predicted $1/4$ suggests that in addition to absorption heating, the intrinsic damping rate increases at higher laser powers, thereby coupling the mechanical oscillator also stronger to its thermal environment. This effect is indeed expected from the temperature-dependence of the mechanical damping rate. 

\section{Optomechanical model}
To fully characterize optomechanical effects in our experiment, we first measure all optical (cavity length $L=58\,\mu$m, finesse $\mathcal{F}= 79,300$, optical injection loss $\eta_\mathrm{opt} = 0.90$, detuning $\Delta = -0.77\times \Omega_\mathrm{cav}$) and mechanical parameters (frequency $\Omega_\mathrm{m}/2\pi=3.5783125$ MHz, quality factor $Q=7\times10^7$) independently as described in the article. 
Fig. \ref{Fig:optdamping}(a) and (b) show the measured and expected mechanical resonance frequency shift and damping rate versus the injected power that were extracted from Brownian motion spectra, ringdown measurements and the optomechanically induced transparency \cite{Weis2010} measurements shown in Fig. \ref{Fig:optdamping}(c). 
Different pairs of input mirror transmission $\mathcal{T}$ and residual cavity loss $\mathcal{L}$ that are consistent with the independently measured finesse are used to model optical spring and damping effects with the standard relations of optomechanics (e.g. from \cite{RMP2014}). 
Both effects are linear versus the injected power to a very good approximation. The best agreement, indicated by yellow lines in Fig. \ref{Fig:optdamping}(a) and (b), is found with the values $\mathcal{T}=39$ ppm and $\mathcal{L}=40$ ppm, which are therefore used for the model in the main paper. An etalon formed between the high-reflectivity coating of the coupling mirror and the substrate's back surface can explain the observed difference between the estimated transmission and its room-temperature value $\mathcal{T}=35$ ppm. Direct transmission measurements with a similar coupling mirror at room temperature have yielded a periodic variation of the transmission with the laser wavelength that is fully consistent with this etalon hypothesis. Together with the environment temperature model introduced in the previous section, the effective mode temperature can thus be modeled independently as explained in the main paper. 
The employed model also includes the effect from quantum backaction, which however only contributes about 10\% of the total force noise for the highest injected optical power. 

\section{Classical laser noise}
Classical laser noise \textbf{sets a lower limit} to the phonon number achievable by optomechanical cooling \cite{Safavi2013}. To rule out classical laser noise in our experiments, we present both intensity and phase noise measurements for our laser source in this Fig. \ref{Fig:lasernoise}. 
For these measurements, our self-built balanced photodetector was first calibrated with quantum shot noise in the balanced configuration. 
The photodiode gain and dark noise level extracted from these measurements are within 10 \% of the expected electronic gains, indicating a good model of the electronic noise sources and a high quantum efficiency of the photodiodes. 
The power dependence of the quantum noise follows a linear relation as a function of laser power. 
For the intensity noise measurement, a range of laser powers was sent onto one of the two photodiodes. 
Calibrated spectra produced for different power levels display good mutual agreement at all frequencies between 0.1 and 10 MHz for the expected quadratic dependence of noise level on laser power. 
For the phase noise measurement, the photodetector was operated in a Mach-Zehnder configuration with an armlength difference below 2 m. 
The relative phase between both laser beams was stabilized at low frequencies (below 10 kHz) such that equal DC-powers were detected on both photodetectors. 
As shown in Fig. \ref{Fig:lasernoise} (b), the classical intensity noise peak of our laser at about 800 kHz is stronly reduced in this measurement, as this scheme is only sensitive to relative phase fluctuations between the two interferometer arms, arising for example from the different vibration of mirrors in the two arms. 
To estimate the frequency noise of our laser source, a 20-m sigle-mode fiber was inserted into the longer arm of the Mach-Zehnder interferometer and the measurement repeated. 
The excess phase noise in this measurement with respect to the previous phase noise measurement is the sum of the laser frequency noise transduced by a 20-m delay line, and the excess phase noise introduced by the single-mode fiber, and thus an upper bound to the frequency noise of our laser source that may affect the measurements presented in the main paper. 
With the highest laser power used in our cooling experiment being 25 $\mu$W, both intensity and phase noise amount to a fraction below $10^{-3}$ of the quantum shot noise of the laser beam in all measurements. 
We therefore conclude that this source of noise is fully masked by the effects of quantum shot noise in the measurements presented in the main paper.